\documentclass[twocolumn,superscriptaddress,aps,preprintnumbers,amsmath,amssymb,prd,nofootinbib,10pt]{revtex4-2}
\usepackage[colorlinks=true,pdfstartview=FitV,breaklinks=true]{hyperref}

\usepackage{physics}
\usepackage{bm}
\usepackage{graphicx}
\usepackage{soul}
\usepackage{amsmath}
\usepackage{mathrsfs}
\usepackage{amssymb}
\usepackage{yfonts}
\usepackage{makecell}
\usepackage{color}
\usepackage{xspace}
\usepackage{url}
\usepackage{verbatim}
\usepackage{mathtools}
\usepackage{upgreek}
\usepackage{units}
\usepackage{amstext}
\usepackage{times}
\usepackage{booktabs}
\usepackage{tabulary}
\usepackage{tabularx}
\usepackage{etoolbox}
\usepackage[version=4]{mhchem}
\usepackage[T1]{fontenc}
\usepackage[dvipsnames,table]{xcolor}
\usepackage{orcidlink}
\usepackage{dcolumn}
\usepackage{bbold}
\usepackage{multirow}
\usepackage{slashed}
\usepackage{colortbl}
\usepackage{lipsum}
\usepackage{cancel}
\usepackage[compat=1.1.0]{tikz-feynman}
\usepackage{tikz,xcolor,hyperref}
\usepackage{adjustbox}

\graphicspath{{Figures/}}

\newcolumntype{Y}{>{\centering\arraybackslash}X}
\hypersetup{urlcolor=BlueViolet,
	    citecolor=Plum,
	    linkcolor=PineGreen}
\usepackage[official]{eurosym}
\definecolor{lightgray}{rgb}{0.9,0.9,0.9}	    
\definecolor{green}{rgb}{0,0.5,0}
\definecolor{red}{rgb}{1,0,0}
\definecolor{blue}{rgb}{0,0,0.5}

\long\def\symbolfootnote[#1]#2{\begingroup%
\def\thefootnote{\fnsymbol{footnote}}\footnotetext[#1]{#2}\footnotemark[#1]\endgroup}

\newcommand\myshade{85}
\colorlet{mylinkcolor}{Blue}
\colorlet{mycitecolor}{Blue}
\colorlet{myurlcolor}{NavyBlue}
\hypersetup{
     linkcolor  = mylinkcolor,
     citecolor  = mycitecolor!\myshade!black,
     urlcolor   = myurlcolor!\myshade!black,
     colorlinks = true,
}

\newcommand{\mbv}{\mbox{$m_{\slashed{B}}$}}
\newcommand{\mubv}{\mbox{$\mu_{\slashed{B}}$}}
\newcommand{\deltabv}{\mbox{$\beta_{\slashed{B}}$}}
\newcommand{\alphabv}{\mbox{$\alpha_{\slashed{B}}$}}
\newcommand{\deltabvp}{\mbox{$\beta_{\slashed{B}}'$}}
\newcommand{\alphabvp}{\mbox{$\alpha_{\slashed{B}}'$}}
\DeclareMathOperator{\MeV}{\text{MeV}}
\DeclareMathOperator{\GeV}{\text{GeV}}
\DeclareMathOperator{\TeV}{\text{TeV}}

\DeclareMathOperator{\Lag}{\mathcal{L}}

\begin{document}

\title{Spontaneous breaking of baryon number, baryogenesis and the bajoron}

\author{Pedro Bittar \orcidlink{0000-0002-3684-5692}}
\email{bittar@if.usp.br}

\author{Gustavo Burdman \orcidlink{0000-0003-4461-2140
}}
\email{gaburdman@usp.br}

\author{Gabriel M. Salla \orcidlink{0000-0002-4283-6637}}
\email{gabriel.massoni.salla@usp.br}

\affiliation{Department of Mathematical Physics,
    Institute of Physics, 
    University of São Paulo, \\
    R. do Matão 1371, São Paulo, 
    SP 05508-090, Brazil
}
\begin{abstract}
We explore the spontaneous breaking of global baryon number for baryogenesis. We introduce a model with three majorana fermions and a complex scalar carrying baryon number charge. After symmetry breaking, the baryon asymmetry is generated below the electroweak scale via the decays of one of the majorana fermions. The main focus of the paper is the phenomenology of the Nambu--Goldstone boson of $U(1)_B$, which we call the bajoron. With small sources of explicit baryon number violation, the bajoron acquires a small mass and is generally very long-lived. The model avoids proton decay, satisfies cosmological constraints, and offers interesting collider phenomenology for the Large Hadron Collider. These long-lived particles tied to baryogenesis strongly support the development of far-detector experiments such as  MATHUSLA, FASER, SHiP, and others.
\end{abstract}

\maketitle

\section{Introduction}\label{sec:intro}
Baryon number ($B$) is the global symmetry carried by quarks that ensures the stability of the proton and controls the dynamics of various hadronic processes. Terrestrial experiments confirm that $B$ is an extremely good symmetry of nature, with no observed processes that violate it\,\cite{Babu:2013jba}. However, as we observe it today, the universe is primarily baryon number asymmetric, with most structures forming visible matter made of particles and not anti-particles. Therefore, it is surprising that $B$ must be both a high-quality symmetry to preserve hadronic baryon number conservation and yet strongly violated to explain the cosmic baryon asymmetry. This discrepancy between these two conflicting observations of our universe has remained a long-standing puzzle in physics.

One possibility is that baryon number violation occurs at very high energies. In the renormalizable Standard Model (SM), $B$ is an accidental symmetry conserved at the classical level. The first non-renormalizable operators that violate $B$ appear at mass dimension six in the effective field theory expansion. Assuming a generic set of $B$-violating operators, 
\begin{equation}
    \mathcal{L}_{d=6}^{\slashed{B}} = \frac{1}{\Lambda_B^2} uude + \dots,
\end{equation}
\noindent the scale $\Lambda_B$ should be roughly of order $\sim 10^{16} \GeV$ to be compatible with bounds from proton decay. These constraints would imply that $B$-violating effects, and potentially baryogenesis, occur far beyond the weak scale of $\sim 100 \GeV$. At the quantum level, baryon number is violated by non-perturbative processes involving electroweak instanton configurations called sphalerons. These also explicitly break global lepton number ($L$), maintaining only the $B-L$ combination anomaly-free. At low temperatures, these processes are not directly observable as they are suppressed by the small probability of tunneling through the associated potential barrier. However, at high temperatures, sphalerons become thermally active and, in principle, could act as sources for the baryon number violation of baryogenesis. Despite these prospects, the SM still fails to meet any criteria for successful baryogenesis\,\cite{Davidson:2008bu,Wagner:2023vqw,Kolb:1990vq}, requiring physics beyond the SM (BSM). Some traditional models that rely on the sphalerons are electroweak baryogenesis\,\cite{Cohen:1993nk} and leptogenesis\,\cite{Davidson:2008bu}. To be compatible with the previously mentioned sources of $B$ and $L$ violation, these models require baryogenesis to occur at very high scales, ranging anywhere from above the weak scale to $10^{16} \GeV$.

While high-scale baryogenesis models provide a compelling picture of ultraviolet (UV) BSM physics, it is still possible that the scale of $B$ violation and baryogenesis is around or below the weak scale. Models of post-sphaleron baryogenesis\,\cite{Dimopoulos:1987rk,Trodden:1999ie,Babu:2006xc,Babu:2006wz,Kohri:2009ka,Allahverdi:2010im,Allahverdi:2022zqr} propose to generate the baryon asymmetry at low scales by introducing a new particle that decays after the sphaleron transition directly to baryon number. These out-of-equilibrium decays violate $CP$ and $B$ to satisfy the Sakharov conditions. Since these decays happen after the sphaleron deactivation, this setup is agnostic on the lepton number content of the universe and does not use the sphaleron processes to transfer the asymmetry from the leptons to the baryons. Most interestingly, low-temperature baryogenesis allows for a rich phenomenology, including prompt searches at the Large Hadron Collider (LHC) for exotic charged scalars, monojet and missing energy signals, long-lived particles (LLPs) and cosmological dark matter generation through cogenesis\,\cite{Curtin:2018mvb,Cui:2012jh,Cui:2013bta,Cui:2014twa,Cui:2016rqt,Ipek:2016bpf,Aitken:2017wie,McKeen:2015cuz,Davoudiasl:2010am,Davoudiasl:2015jja,Arnold:2012sd,Assad:2017iib,Cheung:2013hza,Allahverdi:2010rh,Allahverdi:2013mza,Allahverdi:2017edd,Bittar:2024fau}, to list a few. 

In a companion paper\,\cite{Bittar:2024fau}, we proposed a simplified model approach for addressing the phenomenology of low-temperature baryogenesis, focusing on LLP searches at the High-Luminosity LHC. The model introduces three new majorana fermions $N = N_{1,2,3}$ and a complex scalar $\Phi$, all SM gauge singlets. In this setup, the fermions $N$ decay to quarks to create the baryon asymmetry. Their coupling to quarks arises from some UV dynamics at a scale of $\Lambda \sim 1$ TeV, governed by the following Lagrangian:
\begin{align}
    \Lag \supset \frac{\kappa}{\Lambda^2} N u d d' + \xi\, N \Phi N + h.c. 
    \label{eq:Leff_noflavor}
\end{align}
where $\kappa$ is a flavor matrix controlling the effective operator $Nudd'$, with $u,d^{(')}$ SM quarks, and $\xi$ are the Yukawa couplings of the dark sector needed for $CP$ violation. Without majorana mass terms, global baryon number charge is conserved in the theory. Therefore, for the model presented in Ref.\,\cite{Bittar:2024fau}, a source of explicit breaking of $B$ must be introduced to achieve baryogenesis. This can be done with the term
\begin{equation}
    \mathcal{L}_{\slashed{B}} = - \frac{1}{2} M N N.
    \label{eq:majmass}
\end{equation}

\noindent As discussed in Ref.\,\cite{Bittar:2024fau}, this simple setup is sufficient to generate a baryon asymmetry at low temperatures and to give LLP signatures at various existing or planned LHC detectors.

In this paper, we adapt the model so that the baryon number and $CP$ violations required by baryogenesis occur spontaneously rather than explicitly. Surprisingly, the spontaneous breaking of global baryon number, without  lepton number violation being involved, has been seldom considered in the literature. The concept was introduced in Refs.\,\cite{Barbieri:1981yr,Mohapatra:1982xz}, with a subsequent study focusing on neutron oscillation phenomenology\,\cite{Berezhiani:2015afa}. In the context of spontaneous baryogenesis introduced in Refs.\,\cite{Cohen:1987vi,Cohen:1988kt}, the spontaneous breaking of baryon number and baryogenesis occurs at very high scales, relying on the dynamics of the scalar field to generate a chemical potential for baryon number. Similar models generate the baryon asymmetry from the axion rotation\,\cite{Co:2019wyp,Co:2022aav}. These mechanisms rely on sphaleron dynamics and differ from the out-of-equilibrium decays we propose in our model.

Spontaneous symmetry breaking (SSB) of global baryon number is achieved by assuming that a $B$ charged scalar $\Phi$ acquires a non-zero vacuum expectation value (vev). The main phenomenological consequence is the appearance of an associated pseudo-Nambu-Goldstone Boson (pNGB). We call this pNGB the \textit{bajoron}.\footnote{In Ref.\,\cite{Cohen:1987vi}, the name \textit{Thermion} is introduced and later renamed \textit{Ílion} in Ref.\,\cite{Cohen:1988kt}. In Ref.\,\cite{Berezhiani:2015afa}, the pNGB is referred to as a \textit{baryonic Majoron}, which aligns more closely with the \textit{bajoron} in our model.} In this sense, the bajoron is the baryon number analog of the majoron, which is the pNGB of global lepton number\,\cite{Chikashige:1980ui,Schechter:1981cv,Lazarides:2018aev,Chao:2023ojl}. In fact, it is often assumed that the majoron can break the non-anomalous $B-L$ symmetry instead of just $L$. When this is true, the majoron should take the role of the bajoron, as it is also responsible for breaking baryon number. However, coupling $N$ to baryon number through Eq.\,\eqref{eq:Leff_noflavor}  and simultaneously to lepton number through the right-handed neutrino portal $\overline{L}H N$, leads to rapid proton decay at tree-level\,\cite{Aitken:2017wie,Alonso-Alvarez:2023bat}. Therefore, to avoid proton decay,  the $N$'s of our model cannot mix with neutrinos, and they must be heavier than $\sim 1 \GeV$. For this reason, the bajoron presented here cannot be considered a $B-L$ majoron, and we must choose between having either the baryon or lepton portal of $N$ to the SM neutral operators, but not both. Furthermore, unlike the majoron case, an interesting feature of having SSB of baryon number is that the portal between SM and the dark sector is necessarily non-renormalizable, as stated in Eq.\,\eqref{eq:Leff_noflavor}. This feature makes the dark sector particles feebly interacting and if light long-lived as well.

In this context, we focus on implementing successful baryogenesis and the phenomenological implications of the baryon number charged dark sector composed of the three majorana fermions $N$, the bajoron, and the radial mode of $\Phi$. To ensure that the bajoron is not massless and to make the model phenomenologically viable, $U(1)_B$ has to be broken explicitly. More precisely, we consider two different sources of soft-breaking, parameterized by the scales $\mu_\slashed{B},m_\slashed{B}$. Since the global baryon number symmetry is restored as we take both explicit breaking scales to zero, up to small instanton effects, they are technically natural. Therefore, these terms are stable under quantum corrections and can be much smaller than the spontaneous breaking scale $f_B$, i.e., we consider $\mu_\slashed{B},m_\slashed{B}\ll f_B$. Although not the driver of the baryon asymmetry generation, we anticipate that explicit breaking is imperative to ensure the model is compatible with cosmological constraints. Otherwise, as to be argued in Sec.\,\ref{sec:pheno}, the bajoron becomes stable and would overclose the universe.

The paper is organized as follows. In Sec.\,\ref{sec:model}, we present the model that implements SSB of baryon number, highlighting the predictions for the mass of the bajoron and its couplings to SM states. Then, in Sec.\,\ref{sec:bgen}, we detail the mechanism of baryogenesis and justify, using resonant enhancement of the $CP$ asymmetry, that the correct baryon asymmetry can be obtained. Afterward, Sec.\,\ref{sec:pheno} is dedicated to investigating how cosmological and terrestrial bounds can constrain our model. In particular, these are controlled mainly by the soft-breaking terms, which are constrained to a finite region of parameter space still permitted by current data. We conclude in Sec.\,\ref{sec:conclusion} and provide additional details in the appendices.

\section{Model}\label{sec:model}

This section introduces the model we consider and establishes its relevant features for later discussion. In particular, we describe the mixing of the majorana fermions, the mass generation for the bajoron, and its couplings to the SM.

In the unbroken phase of $U(1)_B$, our model is defined by the SM augmented with three majorana fermions $N_B,\chi_2,\chi_3$ and one complex scalar $\Phi$:
\begin{alignat}{5}
    \nonumber \Lag = &~\Lag_\text{SM}+ \frac{1}{2}\overline{N_B}i ~\slashed{\partial} N_B + \frac{1}{2}\overline{\chi_a}i\slashed{\partial} \chi_a - \frac{1}{2} M_a  \overline{\chi_a^c} \chi_a&&
    \\
    &\nonumber -\frac{\kappa_{B}}{\Lambda^2}(\overline{N^c_B} u_R)({\overline{d_R^c}} d_R) - \xi_{a} \overline{N_B^c} \Phi \chi_a +h.c. &&
    \\
    &\nonumber  
    + |\partial^\mu \Phi|^2 - \lambda_\Phi \Big( |\Phi|^2 - \frac{f_B^2}{2} \Big)^2 &&
    \\
    & + \Delta\mathcal{L}_{\slashed{B}}^{(1)} + \Delta\mathcal{L}_{\slashed{B}}^{(2)} .
    &&
    \label{eq:Leff}
\end{alignat}

\noindent Here, $a=2,3$ are the $\chi$ generation indices, $\kappa_B$ is a matrix of three quark flavor indices and $f_B$ is the scale of SSB of baryon number.
In the second line of the equation above, we have the analogous of Eq.\,\eqref{eq:Leff_noflavor}. In this Lagrangian, a non-trivial baryon number charge can be assigned to the new states and is only broken by the soft terms in the last line of the equation. The associated quantum numbers of the relevant particles are summarized in Tab.\,\ref{tab:QNumbers}. In particular, $N_B$ and $\Phi$ are charged under $U(1)_B$, while $\chi_{2,3}$ are complete singlets. The accidental $U(1)_B$ also implies that only $N_B$ enters in the effective operator of Eq.\,\eqref{eq:Leff_noflavor}, allowing it to interact directly with  quarks. For this effective operator, we assume a particular spinor contraction, indicated by the parentheses. Because the color indices are contracted with a Levi--Civita tensor, $\kappa_B$ is antisymmetric in the $d_R$ flavor indices.\footnote{This particular structure is motivated to avoid proton decay at tree-level. We will go back to this point in Sec.\,\ref{sec:pheno} and refer to App.\,\ref{app:UV} for more details on a UV realization.} For the majorana masses, we assume sizable values compared to the SSB scale, i.e., $M_a\gtrsim f_B$, which we take to be nearly degenerate:
\begin{equation}\label{eq:mass_splitting}
    M_2 = M,\quad  M_3 = M + \delta M, \hspace{0.4cm}\text{ with } \delta M \ll M.
\end{equation}
The mass splitting $\delta M$ needs to be small in order to maximize the $CP$ asymmetry needed for baryogenesis through resonant effects (see Sec.\,\ref{sec:bgen}). The third line of Eq.\,\eqref{eq:Leff} describes the new scalar sector that breaks $U(1)_B$ spontaneously. For simplicity, we consider no quartic interaction between the SM Higgs and $\Phi$. In the last line of the Lagrangian above, we introduce the two soft-breaking terms:
\begin{align}
    &\Delta\mathcal{L}_{\slashed{B}}^{(1)} = \mubv \left( e^{i \alphabv} \Phi + e^{-i\alphabv} \Phi^{\dagger}  \right)|H|^2,
    \label{eq:BV1}
    \\
    &\Delta\mathcal{L}_{\slashed{B}}^{(2)} = -\frac{1}{2}\mbv \overline{N_B^c} N_B,
    \label{eq:BV2}
\end{align}

\noindent where $H$ is the SM Higgs and $\mubv,\mbv$ are real dimensionful parameters that respect $\mbv,\mubv\ll f_B, M$, so that the baryon charge assignment is approximately respected. In the first term, $e^{i\alphabv}$ is a complex phase defined so that $\mubv$ is positive. Similarly, $N_B$ can be re-phased  to make $\mbv$ positive as well. As we will see, the presence of both terms, $\Delta\mathcal{L}_\slashed{B}^{(1,2)}$, allow for the bajoron to acquire a non-vanishing small mass and sizable couplings to the SM. In addition, a suitable choice of $\mbv$ prevents proton decay via $p^+ \rightarrow K^+ N$ by controlling the mass of the lightest eigenstate of the majorana fermions, as described next.

\bgroup
\def\arraystretch{1.2}%
\setlength\tabcolsep{1.6mm}
    \begin{table}
    \centering
        \begin{tabular}{|c|ccccccc|}
        \hline  
        & $Q_L$ & $u_R$ & $d_R$ & $N_B$ & $\chi_{2,3}$ & $\Phi$ & $H$   
        \\ \hline
        $SU(3)_c$ & $3$ & $3$ & $3$ & $-$ & $-$ & $-$ & $-$   
        \\
        $SU(2)_L$ & $2$ & $-$ & $-$ & $-$ & $-$ & $-$ & $2$   
        \\
        $U(1)_Y$  & $1/6$ & $2/3$ & $-1/3$ & $-$ & $-$ & $-$ & $1/2$ 
        \\ \hline
        $U(1)_B$  & $1/3$ & $1/3$ & $1/3$ & $-1$ & 0 & $1$ & $0$ 
        \\ \hline
        \end{tabular}
        \caption{SM Gauge group representations and global baryon number of the particles in the model. $N_B$, $\xi_{2,3}$, and $\Phi$ are the SM-singlets defined in the baryon number unbroken phase.
        }
        \label{tab:QNumbers}
    \end{table}
\egroup

After $\Phi$ gets a vev, it can be rewritten in terms of the radial mode $\phi(x)$ and of the bajoron $b(x)$ fields  as:
\begin{equation}
    \Phi(x) = \frac{1}{\sqrt{2}} e^{ \frac{ib(x)}{f_B}} \Big( f_B + \phi(x)\Big).
    \label{eq:Phi_vev}
\end{equation}
Because of the Yukawa interactions in Eq.\,\eqref{eq:Leff}, $f_B$ contributes to the majorana mass matrix in the form of mixing terms:
\begin{equation}
    -\frac{1}{2} 
    \begin{pmatrix}
        \overline{N_B^ c} \\[3pt]
        \overline{\chi_2^c} \\[3pt]
        \overline{\chi_3^ c} \\
    \end{pmatrix}^T
    \begin{pmatrix}
        \mbv & \sqrt{2}\xi_{2} f_B & \sqrt{2}\xi_{3} f_B
        \\[3pt]
        \sqrt{2}\xi_{2}^* f_B & M  & 0
        \\[3pt]
        \sqrt{2}\xi_{3}^* f_B & 0 & M + \delta M
        \\
    \end{pmatrix}
    \begin{pmatrix}
        N_B \\[3pt]
        \chi_2 \\[3pt]
        \chi_3 \\
    \end{pmatrix}.
    \label{eq:mass_matrix}
\end{equation}
The matrix above can be diagonalized by a unitary matrix $U$, and we label the corresponding mass eigenstates as $N\equiv N_{1,2,3}$. 

As we discuss in Sec.\,\ref{sec:bgen}, a general feature of the model is that the mixing of $\chi_2$ with other states must be small, ensuring a long lifetime for the decay of $N_2$ to quarks. In this limit, the largest mixing occurs between $N_B$ and $\chi_3$, with the mixing angle hierarchy $\theta_{13} \gg \theta_{12}, \theta_{23}$. At zeroth $\delta M$ order, the masses of $N_\alpha$ are given by
\begin{alignat}{5}
    m_{N_1} &\simeq \frac{1}{2} \left(\scalebox{0.9}{$ M+\mbv - \sqrt{ (M-\mbv)^2 + 8 f_B^2 \left(|\xi_2|^2+|\xi_3|^2\right)}  $} \right) ,
    \\
    m_{N_2} &\simeq M ,
    \\
    m_{N_3} &\simeq \frac{1}{2}\left( \scalebox{0.9}{$ M+\mbv + \sqrt{ (M-\mbv)^2 + 8 f_B^2 \left(|\xi_2|^2+|\xi_3|^2\right) } $} \right) .
    \label{eq:masses}
\end{alignat}

\noindent Note that, without the perturbation $\delta M$, $N_2$ does not mix with the other states. This means that the mixing angle $\theta_{12}$ must vanish at zeroth $\delta M$ order. Keeping the linear terms in $\delta M$, the mixing angles are given by
\begin{align}
    & \theta_{23} \simeq \frac{|\xi_2|}{\sqrt{|\xi_2|^2+|\xi_3|^2}},\nonumber
    \\
    & \theta_{12} \simeq  \delta M \frac{ |\xi_2|| \xi_3|}{\sqrt{2} f_B (|\xi_2|^2 + |\xi_3|^2)^{3/2}}\label{eq:fermion_mixing} ,
    \\
    & \theta_{13} \simeq \frac{1}{2}\sin^{-1}\sqrt{\frac{8 f_B^2 \left(|\xi_2|^2+|\xi_3|^2\right)}{(M-m_{\slashed{B}})^2 + 8 f_B^2 \left(|\xi_2|^2+|\xi_3|^2\right)} },\nonumber
\end{align}

\noindent where we discarded terms of order $\sim \delta M \frac{f_B}{M}$ in the expressions for $\theta_{12}$ and omitted the order $\sim \delta M$ contribution to $\theta_{13}$ and $\theta_{23}$. For more details on the diagonalization, we refer the reader to App.\,\ref{app:model_details}.

Let us now discuss the generation of a mass for the bajoron. In principle, we would need to compute the full scalar potential $V(h,\phi,b)$, with $h$ the radial mode of $H$, and then diagonalize the corresponding mass matrix. However, the bajoron mass and the mixing terms must necessarily be proportional to the soft-breaking parameters $\mubv,\mbv$. Hence, working under the approximation $\mubv,\mbv\ll f_B,v$, with  $v$  the Higgs vev, we can neglect the corrections to the bajoron mass coming from mixing with the other scalars. We thus only need to obtain the $b$-dependent term of the potential, $V(b)$. The latter  is determined by the two Lagrangians in Eqs.\,\eqref{eq:BV1} and \eqref{eq:BV2}. After $H$ and $\Phi$ get their vevs, we obtain from Eq.\,\eqref{eq:BV1}:
\begin{equation}
    V^{(1)}(b) = -\mubv f_B v^2 \cos\left( \frac{b}{f_B} + \alphabv \right).
    \label{eq:V1}
\end{equation}
Next, to find the bajoron potential induced by the baryon number violating majorana mass of Eq.\,\eqref{eq:BV2}, we calculate the one-loop effective potential assuming zero momentum external bajoron lines. Rewriting the relevant interactions from Eq.\,\eqref{eq:Leff} using Eq.\,\eqref{eq:Phi_vev} leads to
\begin{align}
    \mathcal{L} \supset - \frac{1}{\sqrt{2}} \left( f_B \xi_a e^{ib/f_B} \overline{N_B^c} \chi_a + f_B \xi_a^\dagger e^{-ib/f_B} \overline{\chi_a^c} N_B \right).
    \label{eq:exp_b}
\end{align}

\begin{figure}[!t]
    \centering
    \includegraphics[scale=1]{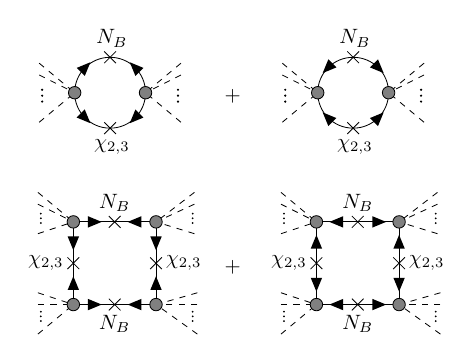}
    \caption{Diagrams contributing to the first two coefficients $a_1$ and $a_2$ of the Coleman--Weinberg potential of the bajoron. The external bajoron legs correspond to the insertion of the $e^{\pm i b/f_B}$ operator, which can be expanded into multiple $b$ fields.}
    \label{fig:VCW_diagrams}
\end{figure}

\noindent  The effective potential is given by the sum of the diagrams of the type shown in Fig.\,\ref{fig:VCW_diagrams}. The external bajoron lines correspond to the exponential interactions in Eq.\,\eqref{eq:exp_b}. To close a loop, we need an even number of $e^{\pm ib/f_B}$ insertions and both the baryon number preserving and violating masses, $M_a$ and $\mbv$, respectively. Evaluating the diagrams results in the following  potential:
\begin{equation}
    \nonumber V^{(2)}(b) = -\sum_{n=1}^\infty c_n \cos\left(\frac{2 n b}{f_B} + n \deltabv\right),
    \label{eq:V2}
\end{equation}

\noindent where the phase $\deltabv$ corresponds to twice the physical phase of the couplings $\xi_2$ and $\xi_3$ in the interaction terms\,\eqref{eq:exp_b}.\footnote{For example, we can take $\xi_a = e^{i\frac{\beta_{\slashed{B}}}{2}} |\xi_a|$ and use Eq.\,\eqref{eq:exp_b} to obtain Eq.\,\eqref{eq:V2}.} In the $\overline{\text{MS}}$ renormalization scheme, the coefficients are given by
\begin{subequations}
\begin{alignat}{5}
    &c_1 = \sum_a\frac{|\xi_a|^2 f_B^2 M_a \mbv}{16\pi^2} \left[ 1 + \log\left(\frac{\mu^2}{M_a^2}\right) \right],
    \\
    &c_2 = \sum_a\frac{|\xi_a|^4 f_B^4}{192\pi^2},
    \\
    &c_r = \sum_a\frac{1}{16\pi^2}\frac{1}{2^n}\frac{(2r-4)!}{(2r-1)!} M_a^4 \left( \frac{|\xi_a|^2f_B^2 \mbv}{M_a^3}\right)^n
    \text{ for } r\geq 3.
\end{alignat}
\end{subequations}

\noindent where $\mu$ is a renormalization scale. The coefficients $c_r$ with $r\geq 3$ are suppressed by powers of the ratios $\mbv/M_a$ and $f_B^2/M_a^2$. Therefore, the leading contribution to the effective potential comes from the coefficient $c_1$. Setting the renormalization scale to $\mu^2=M^2$, the bajoron potential is given by the sum of Eqs.\,\eqref{eq:V1} and \eqref{eq:V2}, $V(b)=V^{(1)}(b)+V^{(2)}(b)$. This potential predicts the following tadpole equation for the bajoron,
\begin{align}\label{eq:bvev}
&\frac{\mubv v^2}{2}\sin(\frac{\langle b \rangle}{f_B}+\alphabv)\\
&\qquad+\frac{f_BM\mbv (|\xi_2|^2+|\xi_3|^2)}{8\pi^2}\sin(\frac{2\langle b \rangle}{f_B}+\deltabv)=0,\nonumber
\end{align}

\noindent that consequently induces a vev $\langle b \rangle$ for the bajoron field. The solution to this tadpole equation is highly non-trivial but simply means that the phases appearing in the expression of the potential $V(b)$ will be redefined as
\begin{equation}
\alphabvp = \frac{\langle b \rangle}{f_B}+\alphabv,\quad \deltabvp = \frac{\langle b \rangle}{f_B}+\deltabv,
\end{equation}
where $\alphabvp,\deltabvp$ are the phases on which observables will depend on. This point will be relevant later in Sec.\,\ref{sec:pheno} and in App.\,\ref{app:model_details} we detail this discussion in more depth. After taking into account the bajoron vev, the mass of the bajoron is extracted to be
\begin{equation}
    m_b^2 \simeq \frac{\mubv v^2}{2 f_B} \cos(\alphabvp) + \left(\frac{|\xi_2|^2 + |\xi_3|^2}{4\pi^2}\right) \mbv M \cos(\deltabvp),
    \label{eq:bmass}
\end{equation}
where we considered only $c_1$ in $V^{(2)}(b)$. As expected, the bajoron mass is proportional to the explicit breaking parameters and the $CP$ phases of the couplings in Eqs.\,\eqref{eq:V1} and \eqref{eq:V2}.

As a brief remark, since the bajoron acquires a vev, an associated symmetry is broken - specifically, $CP$ in this case. A similar argument appears in QCD-axion models when there is an explicit breaking of $U(1)_\text{PQ}$\,\cite{Kallosh:1995hi,Barr:1992qq,Kamionkowski:1992mf,Holman:1992us,DiLuzio:2020wdo}. While for the axion an explicit source of $U(1)_\text{PQ}$ breaking can undermine the solution to the strong $CP$ problem, the bajoron does not face such complications. The reason is that the electroweak vacuum angle, $\theta_\text{EW}$, need not to be small as it does not have any observable effect on the visible sector other than the bajoron-Higgs interactions we describe below.\footnote{Notice that $\theta_\text{EW}$ cannot be rotated away as in the SM because we introduce explicit sources of $U(1)_B$ violation.}

As stressed before, the full scalar potential $V(h,\phi,b)$ contains information on the mixing among the scalars, which is relevant for the phenomenology to be described in Sec.\,\ref{sec:pheno}. From Eq.\,\eqref{eq:BV1} we have
\begin{equation}
    \Delta\mathcal{L}_{\slashed{B}}^{(1)} \supset - \frac{1}{2}\mubv f_B (v+h)^2 \cos\left( \frac{b}{f_B} + \alphabvp \right),
    \label{eq:HCosb}
\end{equation}

\noindent A similar interaction can be found in the backreaction terms of the relaxion mechanism\,\cite{Graham:2015cka,Flacke:2016szy}, for example. The Lagrangian above generates a mixing with the Higgs, given approximately by
\begin{equation}\label{eq:bh_mixing}
    \theta_{hb} \simeq -\frac{\mubv v}{m_h^2} \sin (\alphabvp),
\end{equation}
where $m_h^2$ is the Higgs mass squared. Notice that because the Higgs is $CP$-even while the bajoron is $CP$-odd, the mixing vanishes if the phase $\alphabvp$ is zero. The bajoron thus inherits all couplings of the Higgs to the SM and can be constrained by a wide range of experimental data. The mixing of $\phi$ with the other scalars, $h$ and $b$, is not considered, as it does not significantly impact the phenomenology (see Sec.\,\ref{sec:pheno}).

Finally, after rotating the fields, we can write the Lagrangian for the physical states $N_1, N_2$ and $N_3$, the radial mode $\phi$ and the bajoron $b$,
\begin{alignat}{5}
    \nonumber & \Lag \supset \frac{1}{2} \Big(\overline{N_\alpha}i ~\slashed{\partial} N_\alpha - m_{N_\alpha} \overline{N_\alpha^c} N_\alpha \Big) - \frac{\overline{\kappa}_\alpha}{\Lambda^2}(\overline{N^c_\alpha} u_R)({\overline{d_R^c}} d_R)
    \\
    &\hspace{2mm} \nonumber  - \frac{\overline{\xi}_{\alpha\beta}f_B}{\sqrt{2}} (e^{ib/f_B}-1) \overline{N_\alpha^c} N_\beta - \frac{\overline{\xi}_{\alpha\beta}}{\sqrt{2}} e^{ib/f_B} \phi \overline{N_\alpha^c} N_\beta + h.c. 
    \\ \nonumber 
    &\hspace{2mm} + \frac{1}{2}(\partial^\mu b)^2  ~+~ \frac{1}{f_B} \phi (\partial_\mu b)^2 ~+~ \frac{1}{2 f_B^2} (\partial_\mu \phi)^2 (\partial_\mu b)^2 - V(b)
    \\ 
    &\hspace{2mm}  
    + \frac{1}{2}(\partial^\mu \phi)^2 - \frac{m_\phi^2}{2} \phi^2 - m_\phi \sqrt{\frac{\lambda_\Phi}{2}} \phi^3 -\frac{\lambda_\Phi}{4}\phi^4,
    \label{eq:L_diag}
\end{alignat}

\noindent where we have defined $m_\phi^2 \simeq 2 \lambda_\Phi f_B^2$. Expanding the bajoron exponential, we arrive at a similar Lagrangian as the one used in Ref.\,\cite{Bittar:2024fau}. One obvious difference is the appearance of a pNGB in the spectrum, with its mass determined by Eq.\,\eqref{eq:bmass}. Additionally, the couplings $\overline{\kappa}_\alpha$ and $\overline{\xi}_{\alpha\beta}$ are obtained from the original couplings $\kappa_B, \xi_2$ and $\xi_3$, together with the majorana masses $M$ and $\delta M$ coming from the mixing induced by the $f_B$ terms. Because of this, a hierarchy of couplings is inherited from the mixing pattern discussed above. For example, the decays of $N_2\rightarrow N_1 b$ or to $N_2 \rightarrow udd'$ are suppressed by a small $\bar \xi_{12} \sim \theta_{12} \xi_2$. In general, the new couplings have a complicated dependence on the mixing angles. For the purpose of baryogenesis, the leading contributions to the relevant couplings of the diagonalized Lagrangian in Eq.\,\eqref{eq:L_diag} are given by 
\begin{alignat}{5}
    \label{eq:xi12xi13}
    &\overline{\xi_{12}} \simeq \xi_2 \frac{M \delta M}{2\xi_3^2 f_B^2},
    \hspace{1.2cm}
    &&\overline{\xi_{13}} \simeq \xi_3,
    \\
    \label{eq:k2k3}
    &\overline{\kappa_{2}} \simeq \kappa_B\frac{\delta M \xi_2}{2f_B \xi_3},
    \hspace{1.2cm}
    &&\overline{\kappa_{3}} \simeq \kappa_B \frac{\sqrt{2} \xi_3 f_B}{M}.
\end{alignat}

\begin{figure}[!t]
    \centering
    \includegraphics[scale=1]{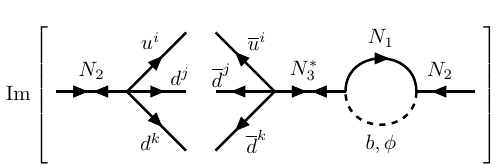}
    \caption{Interference diagrams between the tree-level and one-loop amplitudes that generate the $CP$ asymmetry in the decay of $N_2$ to quarks. The $CP$ asymmetry parameter, $\epsilon_{CP}$, is proportional to the imaginary part of the amplitudes. In the self-energy loop amplitude, $N_3$ is off-shell, and an imaginary part arises from the on-shell intermediate $N_1$ and $b,\phi$ states.}
    \label{fig:CPV}
\end{figure}

\section{Baryogenesis}\label{sec:bgen}

Now that we have defined the model in terms of the physical states $N_\alpha$, $\phi$, and $b$, we can study the dynamics of baryogenesis. The relevant quantity to calculate is the baryon asymmetry parameter, given by the difference of the number of baryons and anti-baryons in the late universe normalized by the entropy density:
\begin{equation}
    Y_{\Delta B}=\frac{n_B - n_{\overline{B}}}{s}.
\end{equation}

\noindent The baryon asymmetry comes from the decay of $N$'s to quarks. As we discuss below, only $N_2$ decays to quarks can violate $CP$. With this consideration, $Y_{\Delta B}$ can be written as
\begin{equation}\label{eq:YB}
    Y_{\Delta B} = Y_{N_2} \,\epsilon_{CP} \,\text{BR}(N_2 \rightarrow udd'),
\end{equation}

\noindent where $Y_{N_2}=n_{N_2}/s$ is the number-to-entropy ratio (yield) of $N_2$ at the time of baryogenesis, $\text{BR}(N_2 \rightarrow udd')$ is the branching ratio of $N_2$ to quarks and $\epsilon_{CP}$ is the $CP$ asymmetry parameter that quantify how much more $N_2$ decays to quarks than anti-quarks:
\begin{equation}
    \epsilon_{CP} = \frac{\Gamma(N_2 \rightarrow udd') - \Gamma(N_2 \rightarrow \overline{udd'})}{\Gamma(N_2 \rightarrow udd') + \Gamma(N_2 \rightarrow \overline{udd'})}.
\end{equation}
We now proceed to compute each of the quantities in Eq.\,\eqref{eq:YB}.

First, the yield of $N_2$ at the time of the asymmetry generation can be estimated at the time the $N_2$ loses thermal contact with the SM bath, after which the SM plasma can no longer wash out the quark-anti-quark asymmetry generated by the decays of $N_2$. To assume that the dark sector was in thermal equilibrium with the SM at some point is the simplest case for the cosmological history of the model, as it does not require any additional production mechanism other than the built-in interactions of the Lagrangian of Eq.\,\eqref{eq:Leff}. The processes that keep $N_2$ in thermal contact with the SM at early times are the annihilation/co-scattering with quarks and the decay of $N_2$ to three quarks. Due to the suppressed interactions with  quarks, it is reasonable to assume that the freeze-out of $N_2$'s interactions occurs when it is still relativistic. To check that, we can calculate the freeze-out temperature $T_\text{FO}$ of $N_2$ by comparing the interaction rate with the Hubble parameter at freeze-out assuming $\Lambda \gg T_\text{FO} \gg m_{N_2}$. We obtain
\begin{equation}\label{eq:TFO}
    T_\text{FO} \simeq 578 \GeV \left( \frac{\Lambda}{1\TeV}\right)^{4/3} \left( \frac{g_*(T_\text{FO})}{106.75}\right)^{1/6} \left| \frac{10^{-6}}{\overline{\kappa}_2}\right|^{2/3}
\end{equation}

\noindent which gives a consistent freeze-out temperature for $m_{N_2} \ll T_\text{FO}$.\footnote{For such high freeze-out temperatures, it might be that $T_\text{FO}\gg f_B$, such that the physical particles would be those in the unbroken phase given in Eq.\,\eqref{eq:Leff}. Our estimates still hold by noticing that the same reactions with similar strength take place for $\chi_{2,3}$ via an intermediate $N_B$ and with an extra $\Phi$.} Then, the yield of $N_2$ at freeze-out is given by the equilibrium distribution 
\begin{equation}
    Y_{N_2}(T_\text{FO}) = \frac{45 \zeta(3)}{2\pi^4} \frac{g_{N_2}}{g_{*s}(T_\text{FO})},
    \label{eq:YieldN2}
\end{equation}

\noindent where $g_{N_2}=2$ is the number of internal degrees of freedom of $N_2$. Here, we denote $g_*(T)$ as the effective number of relativistic degrees of freedom and $g_{*s}(T)$ as the entropy number of degrees of freedom, which, for the temperatures involved, are approximately equal. 

After chemical decoupling from the SM plasma, $N_2$ can still be in thermal contact with the remaining dark sector. Additionally, for energies above the breaking scale, the dark sector particles $N_1,N_3,\Phi$ typically reach thermal equilibrium with the SM. Because thermal equilibrium can be maintained within the dark sector for a longer time, the yield of $N_2$ that sources the baryon asymmetry can replenished as $N_2$ decays to quarks. This effect is more significant the shorter the lifetime of $N_2$ is and the stronger the dark sector interactions are since $N_2$ is kept longer in its equilibrium distribution while decaying to quarks. While this effect can provide an interesting mechanism for enhancing the baryon asymmetry, since the lifetime of $N_2$ is typically large and the dark sector is feebly interacting, it is reasonable to have Eq.\,\eqref{eq:YieldN2} as a conservative estimate for the yield generating the baryon asymmetry. A more precise computation would require solving the complete set of Boltzmann equations and left for future work.

Moving to the branching ratio of $N_2$ to quarks, $N_2$ has two decay channels: $N_2\rightarrow N_1 \phi/b$ and $N_2\rightarrow udd'$. Since both couplings appear due to the mixing induced by the coupling $\xi_2$, the decay of $N_2$ to $N_1 \phi/b$ always dominates over the decay to quarks, which is phase-space and $1/\Lambda^2$ suppressed. Using Eqs.\,\eqref{eq:xi12xi13} and \eqref{eq:k2k3}, the branching to quarks is given by
\begin{align}
    \text{BR}(N_2 \rightarrow udd') \simeq &\left[1 + 32 \pi^2 \left|\frac{\Lambda^2}{m_{N_2}^2}\frac{\overline{\xi}_{12}}{\overline{\kappa}_2}\right|^2 \right]^{-1},
    \label{eq:BrN2toquarks}
    \\
    & \simeq \frac{|\kappa_B \xi_3|^2}{32\pi^2 }\left(\frac{f_B M}{\Lambda^2}\right)^2.
\end{align}

\noindent We see that the branching ratio into quarks is suppressed. Thus,  the resonant enhancement of the $CP$ asymmetry will be a particularly desirable feature in this model, since it will allow for a baryon asymmetry compatible with the observed value, as we discuss now.

The last step is to compute the $CP$ asymmetry $\epsilon_{CP}$. A non-vanishing value for this parameter arises from the interference between tree-level and one-loop contributions to the decay $N\to udd'$. More specifically, this requires a loop amplitude with on-shell intermediate $N\phi/b$ states, where one $N$ mixes with a heavier $N$ flavor. Since the decaying $N$ must be heavier than the $N$ inside the loop but lighter than the one it mixes with, the only flavor capable of generating a $CP$ asymmetry is $N_2$, as $m_{N_1}< m_{N_2}< m_{N_3}$. In addition to the on-shell intermediate states, we also need a non-zero relative phase between the couplings involved in the decay diagrams. The decay diagrams that interfere to generate the $CP$ asymmetry are shown in Fig.\,\ref{fig:CPV}. Due to the small mass splitting among $N_3$ and $N_2$ (see Eq.\,\eqref{eq:mass_splitting}), the $CP$ asymmetry will be enhanced by a resonant effect, similar to what occurs in models of resonant leptogenesis\,\cite{Pilaftsis:2003gt,Pilaftsis:2005rv,Davidson:2008bu}. This assumption requires a slight tuning of the mass splitting parameter $\delta M$ as shown below. The $CP$ asymmetry is extracted from the imaginary part of the interference term in Fig.\,\ref{fig:CPV}:
\begin{equation}
    \epsilon_{CP} = \frac{1}{8\pi} \frac{\Im \left[ \overline{\kappa}_2\overline{\xi}_{12}^\dagger \overline{\xi}_{13} \overline{\kappa}_3\right]}{|\overline{\kappa}_2|^2} \frac{m_{N_2} m_{N_3}\Delta m^2_{32}}{(\Delta m^2_{32})^2 + m_{N_3}^2 \Gamma_{N_3}^2},
    \label{eq:CPpar}
\end{equation}

\noindent where $\Gamma_{N_3}$ is total decay width of $N_3$ and $\Delta m^2_{32} = m_3^2 -m_2^2$. We have included the width of the propagator of $N_3$ to regulate the amplitude for small mass splittings, and a sum over the quark-flavor indices is implied. The largest contribution from the loop occurs when the intermediate states are on-shell, leaving an imaginary contribution from the unitarity cut. This contribution from the loop combines with the real part of the Breit--Wigner propagator of $N_3$ to give the last term shown in Eq.\,\eqref{eq:CPpar}. Since $\overline{\xi}_{13} \simeq \xi_3$ is an $\mathcal{O}(1)$ coupling, the total decay width of $N_3$ is dominated by decays to $N_1$ and a scalar $\phi$ or $b$.
\begin{equation}
    \Gamma_{N_3} \simeq \frac{m_{N_3} |\overline{\xi}_{13}|^2}{8\pi}
\end{equation}

\noindent We can use the expressions in Eqs.\,\eqref{eq:xi12xi13} and \eqref{eq:k2k3} to write Eq.\,\eqref{eq:CPpar} in terms of the parameters of the original interaction basis Lagrangian. If we define the phase of the couplings in the numerator of Eq.\,\eqref{eq:CPpar} as $\varphi$, we have
\begin{equation}
    \epsilon_{CP} = \frac{\Gamma_{N_3}}{m_{N_3}} \frac{m_{N_2}m_{N_3}\Delta m^2_{32}  }{(\Delta m_{32}^2)^2 + m_{N_3}^2 \Gamma_{N_{3}}^2  } \sin\varphi
    \label{eq:CPeps}
\end{equation}

\noindent Notice that, since the couplings in the mass basis all come from the same rotations, the $CP$ asymmetry parameter can be completely rewritten as a function of the masses, the total width of $N_3$ and the phase of the couplings, $\varphi$. A large $CP$ asymmetry is possible if the resonance condition,
\begin{equation}
    m_{N_3} - m_{N_2} \sim \frac{1}{2}\Gamma_{N_3},
    \label{eq:resonance}
\end{equation}

\noindent is satisfied. 
In resonant leptogenesis, the lepton asymmetry is generated at high temperatures, where thermal corrections can modify the resonance condition \eqref{eq:resonance}. In contrast, in our case, baryogenesis takes place much later in the cosmological evolution. As we discuss by the end of the section, the baryon asymmetry is generated at temperatures of $T_{\rm Bgen} \sim 1 \GeV$, making thermal corrections to the resonant condition in Eq.\,\eqref{eq:resonance} negligible.
Assuming $f_B \ll M$ and neglecting terms of order $|\xi_2|^2 \delta M$, we write the mass difference as (see Eqs.\,\eqref{eq:masses} and \eqref{eq:masses_deltaM})
\begin{equation}
    m_{N_3} - m_{N_2} \simeq \delta M + \frac{2(|\xi_2|^2 + |\xi_3|^2)f_B^2}{M}.
\end{equation}

\noindent Therefore, a resonant enhancement of the $CP$ asymmetry happens by tuning the mass splitting $\delta M$ to be
\begin{equation}
    \frac{\delta M}{M} \sim \frac{|\xi_3|}{16\pi}.
\end{equation}

\noindent If the condition above is respected, the $CP$ asymmetry parameters becomes order $\mathcal{O}(1)$, approaching from below the maximal allowed value of Eq.\,\eqref{eq:CPeps}, $\epsilon_{CP}\lesssim 1/2$.

Putting together Eqs.\,\eqref{eq:YieldN2}, \eqref{eq:CPeps} and \eqref{eq:BrN2toquarks}, we can obtain the baryon asymmetry parameter:

\begin{equation}\label{eq:YB_final}
    \frac{Y_{\Delta B}}{Y_{\Delta B}^{\rm exp}} \simeq \left(\frac{1 \TeV}{\Lambda} \right)^4\left(\frac{M}{800 \GeV} \frac{f_B}{100 \GeV} \frac{|\kappa_B|}{0.1}\frac{|\xi_3|}{0.4}\right)^2,
\end{equation}

\vspace{0.5cm}

\noindent where $Y_{\Delta B}^{\rm exp} \approx 8.7\times 10^{-11}$ is the observed central value of the baryon asymmetry measured by Planck\,\cite{Planck:2018vyg}. From Eq.\,\eqref{eq:YB_final}, it thus becomes evident that baryogenesis can be successful for low SSB scales, possibly much below the electroweak scale. 
Finally, to determine the scale of baryogenesis, we can calculate the lifetime of $N_2$ in the quark channel.
\begin{align} \nonumber
    \tau_{N_2\rightarrow udd'} \simeq & 10^{-6}\text{s} \left( \frac{\Lambda}{1\TeV}\right)^4 \left( \frac{f_B}{10^2 \GeV} \right)^2 
    \\
    &\qquad\times \left(\frac{700 \GeV}{M} \right)^7 \left(\frac{0.1}{\kappa_B} \right)^2 \left( \frac{10^{-6}}{\xi_2}\right)^2
\end{align}

\noindent This lifetime points to baryogenesis happening while the universe was around $T_{\rm Bgen} \sim 1 \GeV$, which is much later than the scale of symmetry breaking.

\section{Constraints and phenomenology}\label{sec:pheno}

Having described how baryogenesis works, we can now discuss the corresponding bounds on the parameter space. There are two different regions to constraint: the one of the parameters $M,f_B,\kappa_B,\xi_3$ that control the baryon asymmetry in Eq.\,\eqref{eq:YB_final}, and the soft-breaking parameters $\mbv,\mubv$ that are related to the phenomenology of the bajoron.

For the parameters relevant to baryogenesis in Eq.\,\eqref{eq:YB_final}, while the Yukawa interactions $\xi_{3}$ is essentially unconstrained, 
$\kappa_B$ can give signatures in flavor physics. For instance, we have bounds from $K^0\bar K^0$ and $B^0\bar B^0$ mixing at one loop. These bounds impose $\bar\kappa_{3,2}\lesssim \order{10^{-2}}$\,\cite{Bittar:2024fau,Giudice:2011ak,Han:2010rf,Baldes:2011mh,Pascual-Dias:2020hxo,Han:2023djl} for the scale $\Lambda\sim 1 \TeV$. Both kaon and $B$-meson oscillation happen at the one-loop level in our setup because $\kappa_B$ is antisymmetric in the down-quark indices (see Sec.\,\ref{sec:model} and App.\,\ref{app:UV}). Further constraints on $\kappa_B$ can be obtained from $\Lambda^0\bar \Lambda^0$ oscillation. For this process, the mixing occurs at tree-level by the exchange of $N$. Notice that due to the antisymmetric structure of the $d_R$ couplings, only baryons made of two different flavors of down quark can oscillate to their antiparticle states at tree-level. This means that there is no neutron oscillation at tree level, and the only oscillating baryons are the $\Lambda^0$'s and, in principle, $\Lambda_b^0, \Xi_b^0$. Using the results from the BESIII collaboration\,\cite{BESIII:2023tge}, we have that\footnote{For this estimate, we assume that $m_{N_1}$ is sufficiently separated from the $\Lambda^0$ mass. Since $\Lambda^0$ has a very small width, this does not impose significant constraints on the parameter space.}
\begin{equation}\label{eq:BESIII}
|\kappa_B| \lesssim 10^{-1}\left(\frac{m_{N_1}}{5~\text{GeV}}\right)^{1/2}\left(\frac{\Lambda}{1~\text{TeV}}\right)^2\quad (\Lambda^0\bar \Lambda^0~\text{oscillation}),
\end{equation}
which can be improved by at least an order of magnitude in the future\,\cite{Achasov:2023gey}. 

Moreover, one can also produce $N$'s directly at the LHC and, given that they can be long-lived thanks to their feeble interactions, the might result in displaced vertex signals at future facilities at the LHC such as, for instance,  MATHUSLA\,\cite{Curtin:2018mvb,Curtin:2023skh}. We refer the reader to our companion paper\,\cite{Bittar:2024fau} for more details on this possibility. Considering the bounds just described, we therefore see that there is a lot of free parameter space  allowing for successful baryogenesis and that, at the same time, can be probed in more depth in the near future.

\begin{figure}[t!]
    \centering
    \includegraphics[width=1\linewidth]{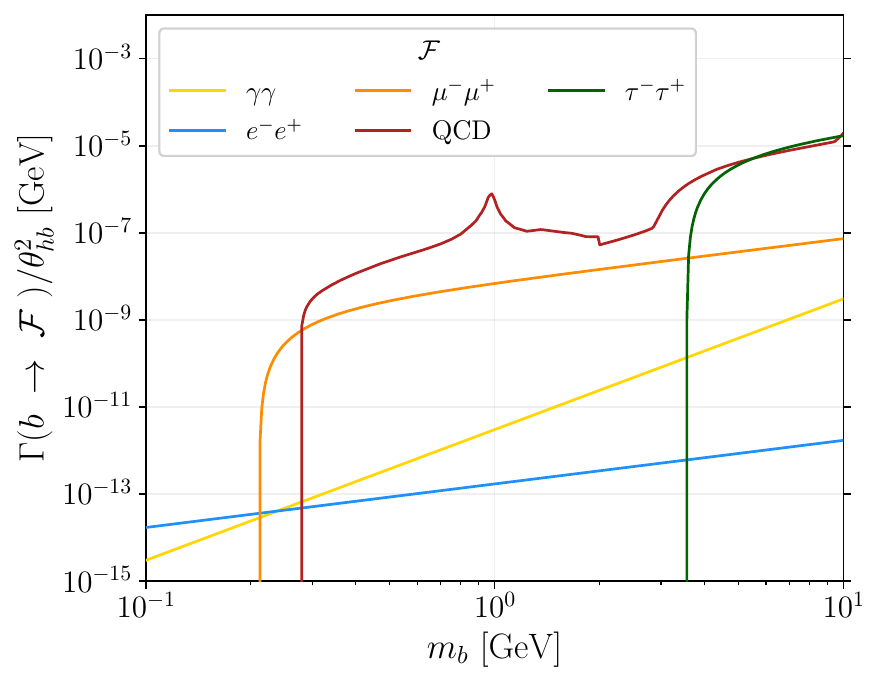}
    \caption{Partial widths of the bajoron, normalized by the mixing angle squared $\theta_{hb}^2$, as a function of its mass. For the QCD width we use the results of Ref.\,\cite{Winkler:2018qyg} up to $m_b=2$ GeV and after this value the perturbative results of decays to quarks.}
    \label{fig:widths}
\end{figure}

We  now turn to the phenomenology of the pNGB of $U(1)_B$, the bajoron. Let us first consider the cosmological history in its presence. As described in Sec.\,\ref{sec:bgen}, before SSB, we have considered that the majorana fermions and $\Phi$ have reached thermal equilibrium with the SM at some point. SSB happens as the temperature lowers to $T\simeq f_B$, and $\Phi$ is replaced by the radial mode $\phi$ and the bajoron $b$. For $m_\phi\lesssim f_B$ and $f_B<v$, decay and inverse decay processes $\phi\leftrightarrow bb$ are extremely efficient and quickly thermalize the bajoron. Since the bajoron is, in general, the lightest particle of the dark sector and its interactions with the other particles are feeble, it will not stay in thermal contact for long, and its final abundance will be given by the relativistic equilibrium yield, $Y_b\simeq 10^{-2}$. Considering $m_b\sim\order{\text{MeV}-\text{GeV}}$, for a stable bajoron\footnote{In the absence of any soft-breaking, the bajoron cannot decay to SM particles and is thus stable. This is easily seen in the unbroken phase in Eq.\,\eqref{eq:Leff}, as the Yukawa interactions are always off-diagonal 
in $N_B,\chi_{2,3}$, while only $N_B$ couples to quarks.} such a yield would produce  too large an energy density, which would overclose the universe. In order to avoid this problem, the bajoron must have a way to decay. The soft-breaking term in Eq.\,\eqref{eq:BV1} predicts a mixing with the Higgs through the angle in Eq.\,\eqref{eq:bh_mixing}, and so the bajoron inherits the same decay channels of the Higgs. The partial widths of the bajoron are given in Fig.\,\ref{fig:widths}. For masses $m_b\lesssim 2$ GeV, the hadronic width is given by the results of Ref.\,\cite{Winkler:2018qyg}. In addition, the bajoron must decay faster than roughly 1 second, otherwise, its decays to SM particles could disrupt Big-Bang Nucleosynthesis (BBN). Then, we constrain its life-time $\tau_b$ as
\begin{equation}\label{eq:BBN}
\tau_b\lesssim 1~\text{s}\quad (\text{BBN}).
\end{equation}
This constraint limits the mass $m_b$ and the mixing angle $\theta_{hb}$ in Eq.\,\eqref{eq:bh_mixing}. A similar constraint was derived in Ref.\,\cite{Fradette:2017sdd}.

Another immediate consequence of the mixing between the bajoron and the Higgs is that the former, much like a dark scalar\,\cite{Silveira:1985rk,Patt:2006fw,OConnell:2006rsp}, can be searched for at various collider experiments. These include analyses that look for exotic meson decays\,\cite{LHCb:2015nkv,LHCb:2016awg,Winkler:2018qyg,Gorbunov:2021ccu,NA62:2021zjw,Belle:2021rcl}, LEP constraints\,\cite{L3:1996ome} and future far detector facilities at the LHC\,\cite{Curtin:2018mvb,Curtin:2023skh,Feng:2022inv,Aielli:2022awh,Bauer:2019vqk}. Many projections were made for future LHC facilities\,\cite{Beacham:2019nyx} that look for displaced vertices, i.e. LLPs. We recast all these limits to our scenario. Here we completely neglect the radial mode $\phi$, which, in principle, could contribute to the production of bajorons at high-energy experiments. We leave a more refined analysis for future work.

Depending on the mass of the light $N_1$, the model can result in proton decay at tree-level. More precisely, proton decay happens by $p^+\to K^+N_1$ via the effective interaction of Eq.\,\eqref{eq:Leff}. Given that $N_1$ is typically much lighter than $M$ and $f_B$, avoiding bounds from proton decay implies
\begin{equation}\label{eq:proton_decay}
m_{N_1}>m_{p^+}-m_{K^+}\quad\text{(proton~decay)},
\end{equation}
which depends a lot on the soft-breaking parameter $\mbv$ (see Eq.\,\eqref{eq:masses}).

There are also two theoretical constraints. First, for our whole picture to be consistent, we must have $\mubv,\mbv\ll f_B$, that is, the amount of soft-breaking is small, and it makes sense to work with spontaneously broken $U(1)_B$. Second, while $m_b$ is predicted as a function of the breaking parameters in Eq.\,\eqref{eq:bmass}, we expect it to be parametrically smaller than the breaking scale $f_B$, otherwise the bajoron could not be considered as a pNGB. Hence, we must impose that $m_b\ll f_B$.

\begin{figure}[t!]
    \centering
    \includegraphics[width=1\linewidth]{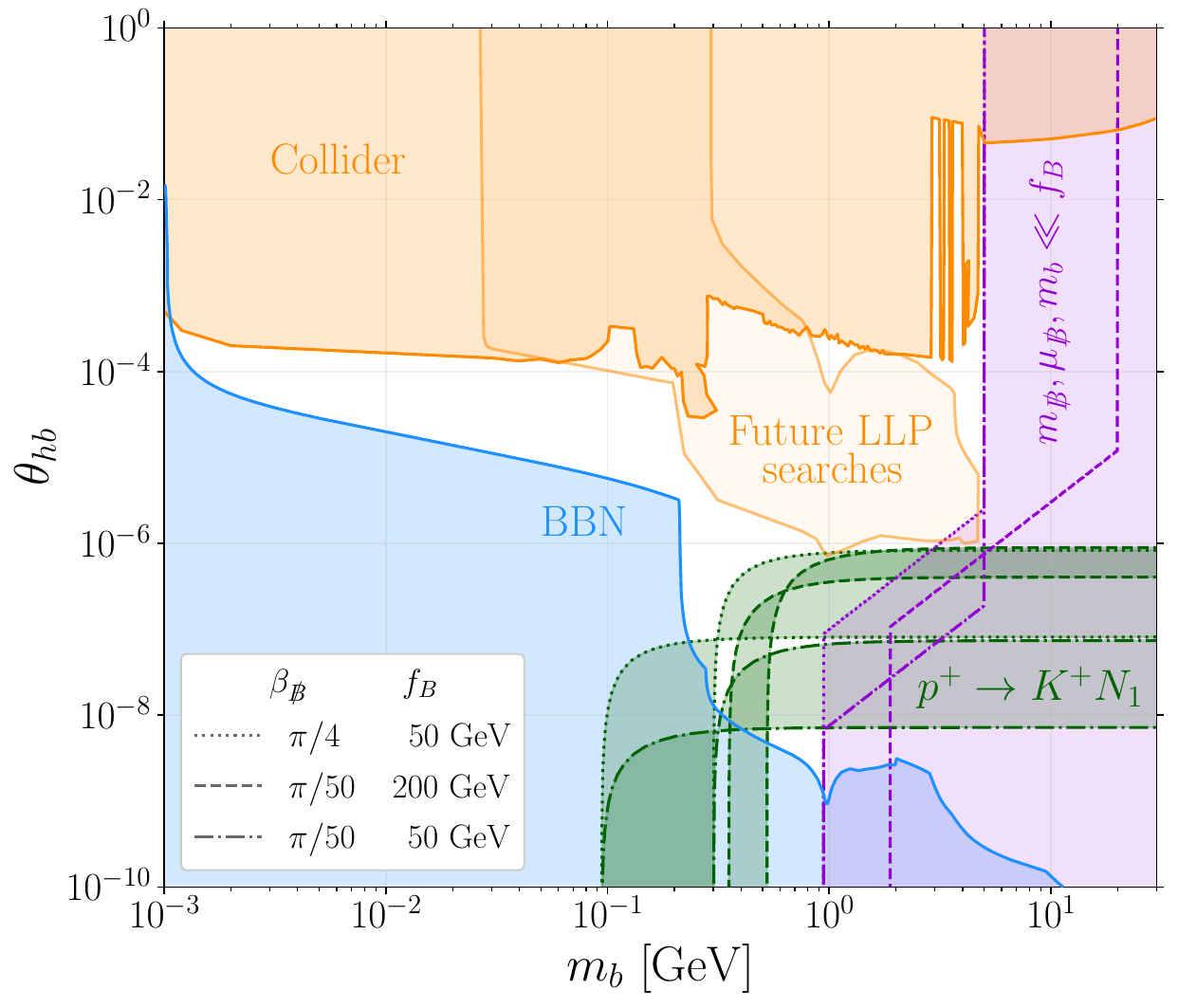}
    \caption{Excluded/Disfavored (colored) regions as a function of the bajoron mass $m_b$ and its mixing angle with the Higgs $\theta_{hb}$ for the benchmark values in Eq.\,\eqref{eq:BM}, with $M=700$ GeV, $|\xi_2|^2\ll|\xi_3|^2=10^{-2}$ and $\alphabv=0$ fixed. The limits obtained for different benchmark points are denoted by dashed, dot-dashed, and dotted lines, while solid lines are for the ones independent of the benchmark chosen. The orange region refers to collider constraints\,\cite{LHCb:2015nkv,LHCb:2016awg,Winkler:2018qyg,Gorbunov:2021ccu,NA62:2021zjw,Belle:2021rcl,L3:1996ome}, lighter orange to future LLP searches\,\cite{Beacham:2019nyx}, blue to the BBN bound in Eq.\,\eqref{eq:BBN}, purple to when $\mbv,\mbv,m_b\ll f_B$ starts becoming invalid and green to where proton decay $p^+\to K^+N_1$ is allowed.}
    \label{fig:exclusion}
\end{figure}

We can now combine everything and determine the parameter space region excluded/disfavored by experimental data. To investigate how these regions vary with the choice of parameters, we define three benchmark (BM) points that can generate the correct baryon asymmetry in Eq.\,\eqref{eq:YB_final} while respecting Eq.\,\eqref{eq:BESIII} and bounds on $K^0\bar K^0$ and $B^0\bar B^0$ oscillation:
\begin{align}\label{eq:BM}
&\text{BM1}:~\deltabv = \frac{\pi}{4}, && f_B=50~\text{GeV},\nonumber\\
&\text{BM2}:~\deltabv = \frac{\pi}{50}, && f_B=200~\text{GeV},\\
&\text{BM3}:~\deltabv = \frac{\pi}{50}, && f_B=50~\text{GeV},\nonumber
\end{align}
holding $\alphabv=0$, $M=700$ GeV and $|\xi_2|^2\ll |\xi_3|^2=10^{-2}$ fixed. Notice that we can, without loss of generality, take either $\alphabv$ or $\deltabv$ to zero, as we can redefine the fields such to to absorb the phases in the couplings; the phase $\alphabv$ in Eq.\,\eqref{eq:BV1} can be re-absorbed in $\xi_{2,3}$ by a redefinition of $\Phi$. In Fig.\,\ref{fig:exclusion}, we show our results in the parameter space $m_b\times \theta_{hb}$  for the three BM points above, where colored regions mean excluded/disfavoured. In orange we have the bounds coming from Refs.\,\,\cite{LHCb:2015nkv,LHCb:2016awg,Winkler:2018qyg,Gorbunov:2021ccu,NA62:2021zjw,Belle:2021rcl,L3:1996ome}, lighter orange denote the future LLP searches\,\cite{Beacham:2019nyx} and in blue the BBN constraint from Eq.\,\eqref{eq:BBN}. These limits are equal for all BM points since the respective observables depend only on the mass $m_b$ and the mixing angle $\theta_{hb}$. In addition, we have the purple regions obtained from the theoretical bounds $\mbv,\mubv,m_b\ll f_B$. To determine them, we do the following: for each BM point, we scan over $\mubv$ and $\mbv$ that violate $\mbv,\mubv \ll f_B$, taking, say, $\mbv/f_B,\mubv/f_B\geq 10^{-1}$ as reference, and afterwards compute $m_b$ and $\theta_{hb}$ according to Eqs.\,\eqref{eq:bmass} and \eqref{eq:bh_mixing}, respectively. This procedure involves solving numerically the tadpole equation\,\eqref{eq:bvev} for the bajoron in order to compute $\alphabvp$ and $\deltabvp$.

The behavior of the purple curves can be readily understood. Consider the limit $\mubv\to 0$, for which $2\langle b \rangle/f_B \to -\deltabv$ and $m_b^2 \to M\mbv(|\xi_2|^2+|\xi_3|^2)/(4\pi^2)$. Hence, the mass $m_b$ becomes independent of $\mubv$, and consequently of $\theta_{hb}$, leading to the left-most vertical lines. In the opposite limit $\mbv\to 0$, the tadpole equation is instead solved by
\begin{equation}
\langle b \rangle/f_B\simeq -\frac{f_BM \mbv(|\xi_2|^2+|\xi_3|^2)}{8\pi^2 v^2\mubv}\sin(\deltabv),
\end{equation}
Thus, as $\mubv$ becomes much larger than $\mbv$, it can be shown that $\theta_{hb}/m_b^2\sim f_B\sin(\deltabv) $, giving rise to the intermediate crescent behavior of the purple curves (see Eq.\,\eqref{eq:theta_mass_scaling}). The point at which this transition happens depends, of course, on $\sim f_B \sin\deltabv$, in agreement with Eq.\,\eqref{eq:bvev}. Finally, the right-most vertical lines mark where $m_b/f_B\ll 1$ starts being violated. In App.\,\ref{app:model_details}, we elaborate further on these points.

Lastly, the green regions in Fig.\,\ref{fig:exclusion} refer to those where proton decay at tree-level is allowed (see Eqs.\,\eqref{eq:masses} and \eqref{eq:proton_decay}). They are computed in a similar fashion as the purple regions: we scan over the soft-breaking parameters, compute $\theta_{hb},m_b$ and $m_{N_1}$ and check whether or not Eq.\,\eqref{eq:proton_decay} is violated. Their dependence with $m_b$ and $\theta_{hb}$ is similar to the one  discussed above for the purple regions.

On the one hand, we observe that collider and BBN bounds, which apply to any scalar mixing with the Higgs, can be directly mapped onto the $m_b\times \theta_{hb}$ parameter space, as they depend solely on the physical mass and mixing angle of the bajoron. On the other hand, constraints explicitly related to our model, namely $\mubv,\mbv,m_b\ll f_B$, and proton decay, are more involved and restrict different parts of the parameter space. In particular, they depend manifestly on the choice of the parameters, evidenced by the distinct BM points of Eq.\,\eqref{eq:BM}. An interesting feature of Fig.\,\ref{fig:exclusion} is that the area of parameter space still not excluded/disfavoured (not colored) is finite, i.e., it is constrained from all sides. This means that (near) future experiments, such as those depicted in Fig.\,\ref{fig:exclusion} for LLP searches, will likely fill all the remaining gaps, thus completely constraining the bajoron parameter space.

\section{Conclusions}\label{sec:conclusion}

In this paper we have considered  baryon number symmetry, $U(1)_B$, as spontaneously broken through the vacuum expectation value of a baryon-charged scalar field. In addition to the complex scalar field, we extend the SM by including three flavors of the Majorana fermions $N_B,\chi_2, \chi_3$, out of which $N_B$ carries baryon number and can interact with quarks. In the broken phase of the model, these majorana fermions mix to become the mass eigenstates $N_1,N_2,N_3$, which can all decay into quarks. This setup provides a mechanism to generate the baryon asymmetry of the universe at temperatures below the electroweak scale of $\sim 100 \GeV$. Spontaneous baryogenesis occurs in the model through the $CP$-violating, out-of-equilibrium decays of the intermediate mass majorana fermion $N_2$. An important feature of the model is the resonant enhancement of $CP$ violation, achieved by tuning the mass splitting between $N_2$ and $N_3$. This leads to a significant baryon asymmetry that is consistent with the observed value measured by the Planck telescope.

The central phenomenological interest of our paper is the appearance of the pseudo-Nambu--Goldstone boson of baryon number, the \textit{bajoron}. In our model the bajoron plays an analogous role to the majoron, the pNGB of global lepton number that appears in models of BSM neutrino masses. Differently from the majoron, in the absence of explicit $U(1)_B$ breaking terms, the bajoron only has non-renormalizable portal interactions to the SM. This fact makes it very feebly interacting and a naturally long-lived particle. 

To make the model cosmologically viable, explicit baryon number violation is introduced through soft-breaking terms, parameterized by $\mbv$ and $\mubv$ (Eqs.\,\eqref{eq:BV1} and \eqref{eq:BV2}). These terms ensure that the bajoron acquires a small mass and decays early enough to avoid overclosing the universe or disrupting Big-Bang Nucleosynthesis. We calculate the bajoron potential that is induced by the soft-breaking terms and derive an expression for the bajoron mass, which generally assumes values between $1\MeV$ to $5\GeV$ (Eq.\,\eqref{eq:bmass}). The bajoron decays to the SM are controlled by its mixing with the Higgs, Eq.\,\eqref{eq:bh_mixing}, and subsequent decays to light SM degrees of freedom. We show the partial widths of the contributing channels in Fig.\,\ref{fig:widths}. 

Finally, in Fig.\,\ref{fig:exclusion}, we present the bounds as an exclusion region in the $m_b \times \theta_{hb}$ plane for different values of the $CP$ phases of the couplings and scales of SSB of baryon number, $f_B$. The allowed region arises from simultaneously requiring decays before Big-Bang Nucleosynthesis, avoiding tree-level proton decay induced by light $N_1$'s, theoretical considerations on the soft-breaking parameters being much smaller than $f_B$ and avoiding regions that are excluded by  present collider experiments. In particular, the collider constraints come from exotic meson decays producing bajorons and electroweak precision data from LEP. Most interestingly, the allowed region greatly overlaps with the projected sensitivity region for far detector facilities at the LHC. These are MATHUSLA, SHiP, CODEX-b, FASER2, and NA62, and some regions coming from the LHCb at run3. Notably, upcoming experiments are expected to explore the remaining parameter space, providing a comprehensive test of our proposed model. Therefore, the baryogenesis scenario we presented in this paper makes a strong physics case in favor of the construction and continuity of the long-lived particle program. In particular, the forward and transverse far detectors mentioned above should greatly increase the reach for new physics and shed light on the nature of the baryon number symmetry and baryogenesis mechanism of our universe. 

\begin{acknowledgments}
We thank Ana Luisa Foguel and Lucas M. D. Ramos for valuable discussions. Additionally, we acknowledge the financial support of "Fundação de Amparo à Pesquisa do Estado de São Paulo" (FAPESP) grant numbers 2019/04837-9, 2020/14713-2 and "Fundação Coordenação de Aperfeiçoamento de Pessoal de Nível Superior" (CAPES) grant number 88887.816450/2023-00.
This project has received funding /support from the European Union’s Horizon 2020 research and innovation programme under the Marie Skłodowska -Curie grant agreement No 860881-HIDDeN.
\end{acknowledgments}

\appendix

\section{More details on the model}\label{app:model_details}

In this appendix we give some more details on the model we have defined in Sec.\,\ref{sec:model}. More precisely, we define more carefully the majorana fermion mixing and present more equations for the scalar sector.

The majorana mass matrix for $(N_B,\chi_2,\chi_3)$ after the SSB is defined by Eq.\,\eqref{eq:mass_matrix}:
\begin{equation}
    M_{N_B\chi}= 
    \begin{pmatrix}
        \mbv & \sqrt{2}\xi_{2} f_B & \sqrt{2}\xi_{3} f_B
        \\[3pt]
        \sqrt{2}\xi_{2}^* f_B & M  & 0
        \\[3pt]
        \sqrt{2}\xi_{3}^* f_B & 0 & M + \delta M
        \\
    \end{pmatrix}.
\end{equation}
In the limit of vanishing mass splitting $\delta M \to 0$, the mass eigenvalues become
\begin{align}
&m_{N_{1,3}}(\delta M=0)\nonumber\\
&= \frac{M+\mbv}{2} \mp \sqrt{\left(\frac{M-\mbv}{2}\right)^2+2f_B^2(|\xi_2|^2+|\xi_3|^2)},\\
&m_{N_2}(\delta M=0) = M,
\end{align}
meaning that the mass eigenstate $N_2$ coincides with the $\chi_2$ state of the unbroken phase. In Sec.\,\ref{sec:bgen} we need the splitting between the mass eigenstates $m_{N_2}$ and $m_{N_3}$, which to leading order in the small parameters is
\begin{align}\label{eq:masses_deltaM}
    &m_{N_1}(\delta M^{(1)}) = \delta M \frac{4 f_B^2 |\xi_3|^2}{\Delta} \frac{1}{M - \mbv + \Delta},
    \\
    &m_{N_2}(\delta M^{(1)}) = \delta M\frac{|\xi_2|^2}{|\xi_2|^2 + |\xi_3|^2},
    \\
    &m_{N_3}(\delta M^{(1)}) =  \frac{4\delta M f_B^2 |\xi_3|^2}{8f_B^2 (|\xi_2|^2 +|\xi_3|^2) + (M - \mbv) \Delta},
\end{align}
with
\begin{equation}
\Delta\equiv \sqrt{(M-\mbv)^2 + 8f_B^2 (|\xi_2|^2 + |\xi_3|^2)}.
\end{equation}
The masses $m_{N_{1,2,3}}$ can turn out to be negative. In such a case, it is just a matter of re-phasing the fields by $e^{i\pi/2}$ and re-absorb it in some of the couplings. This procedure does not affect any of our conclusion. The rotation matrix $U$ that diagonalizes $M_{N_B\chi}$ is defined by
\begin{equation}
    U= R_3(\theta_{12}) R_2(\theta_{13},\delta) R_1(\theta_{23})
\end{equation}
as a function of mixing angles $\theta_{12},\theta_{23},\theta_{13}$ and a phase $\delta$, where $R_i$ are the usual rotation matrices along each axis. To first order in $\delta M$, the angles are given by Eq.\,\eqref{eq:fermion_mixing}, making explicit that $\theta_{12}\to 0$ as $\delta M\to 0$.

Turning to the bajoron potential, the loop integrals that give rise to the coefficients $c_n$ are given by
\begin{equation}
    c_n = 2\hspace{-1mm}\sum_{a=2,3}\hspace{-1mm} \left(\frac{|\xi_a| f_B}{\sqrt{2}}\right)^{2n} \hspace{-1mm}\int \frac{d^4k}{(2\pi)^4} \left(\hspace{-0.5mm}\frac{M_a}{k^2- M_a^2} \frac{\mbv}{k^2- \mbv^2}\hspace{-0.5mm} \right)^n.
    \label{eq:c_n}
\end{equation}
\noindent Notice that the equation above for $n\geq 2$ only includes diagonal terms, i.e. loops that contain just a single species of $\chi$. In general, we can also have mixed loops for $n\geq 2$, nevertheless they are irrelevant to our discussion. All $c_n$ except for $n=1$ turn out to be finite. The coefficient $c_1$ can be renormalized using the $\overline{\text{MS}}$ scheme, and taking the renormalization scale to $M$ eliminates the log dependence.

Let us now discuss the scalar potential. The full potential in the unbroken phase $V(H,\Phi)$ is given by
\begin{equation}
V(H,\Phi) = -\mu^2 |H|^2 + \lambda|H|^4 -\mu_\Phi^2|\Phi|^2+\lambda_\Phi |\Phi|^4 - \Delta\mathcal{L}^{(1)}_\slashed{B},
\end{equation}
where $\mu^2,\mu_\Phi^2$ and $\lambda^2,\lambda_\Phi^2$ are the respective mass parameters and quartic couplings, respectively, and $\Delta\mathcal{L}^{(1)}_\slashed{B}$ is defined in Eq.\,\eqref{eq:BV1}. 
After the SSB of both SM and $U(1)_B$ groups, we can now expand $H$ and $\Phi$ around their vevs, $v$ and $f_B$. As stressed in Sec.\,\ref{sec:model}, a Coleman--Weinberg potential\,\eqref{eq:V2} generated at one-loop is also relevant for the bajoron. Taking it into account, the tadpole equations that define the vevs are
\begin{align}\label{eq:tadpole_system}
& v^2\lambda - \mu^2 -\mubv f_B  \cos(\frac{\langle b \rangle}{f_B}+\alphabv )=0,\\
&f_B^3\lambda_\Phi - f_B\mu_\Phi^2 - \frac{v^2\mubv}{2}\cos(\frac{\langle b \rangle}{f_B}+\alphabv)=0,\\
&\frac{\mubv v^2}{2}\sin(\frac{\langle b \rangle}{f_B}+\alphabv)\label{eq:b_tadpole}\\
&\qquad+\frac{f_BM\mbv (|\xi_2|^2+|\xi_3|^2)}{8\pi^2}\sin(\frac{2\langle b \rangle}{f_B}+\deltabv)=0.\nonumber
\end{align}
Because of the bajoron vev $\langle b \rangle$ induced by the phases $\alphabv,\deltabv$, the system above is not exactly solvable in general. In spite of this, we can work in the approximation that $\mubv,\mbv\ll f_B,v$ and expand to first order in the soft-breaking terms. Doing so we obtain
\begin{align}
v^2 \simeq \frac{\mu^2}{\lambda},\quad f_B^2 \simeq \frac{\mu_\Phi^2}{\lambda_\Phi},
\end{align}
which lead to the usual expressions for the radial modes. The bajoron tadpole is instead much more involved. The solution of $\langle b \rangle$ is crucial, since the physical phases which the mixing angle $\theta_{hb}$\,\eqref{eq:bh_mixing} and the bajoron mass $m_b^2\,\eqref{eq:bmass}$ depend on are
\begin{align}
\alphabvp = \alphabv + \frac{\langle b \rangle}{f_B},\quad \deltabvp = \deltabv + \frac{2\langle b \rangle}{f_B}.
\end{align}
Notice that it is possible to fix $\alphabv=0$, since we can perform a field redefinition on $\Phi$ and re-absorb $\alphabv$ in the definition of $\xi_{2,3}$ and, consequently, of $\deltabv$.

As argued in Sec.\,\ref{sec:pheno}, it is necessary for the bajoron to mix with the Higgs in order for it not to overclose the universe. A non-vanishing mixing angle $\theta_{hb}$ can only be obtained if both soft-breaking parameters are non-zero $\mubv,\mbv\neq 0$. To see this, consider first the limit $\mbv\to 0$, for which Eq.\,\eqref{eq:b_tadpole} becomes
\begin{equation}
\mbv\to 0:\quad \frac{\mubv v^2}{2}\sin(\frac{\langle b \rangle}{f_B}) = 0\Rightarrow \frac{\langle b \rangle}{f_B} = 0\Rightarrow \theta_{hb}\to 0,
\end{equation}
and therefore the mixing is zero. Therefore, both soft-breaking parameters are vital to guarantee a phenomenologically viable model. When the term proportional to $\mubv$ in Eq.\,\eqref{eq:b_tadpole} dominates over the one proportional to $\mbv$, but $\mbv\neq 0$, the bajoron vev is given to first order by
\begin{equation}
\langle b \rangle/f_B\simeq -\frac{f_BM \mbv(|\xi_2|^2+|\xi_3|^2)}{8\pi^2 v^2\mubv}\sin(\deltabv).
\end{equation}
Since in this case $|\langle b \rangle /f_B|\ll 1$, we can expand the expression for the mixing angle and for the mass $m_b^2$, in Eqs.\,\eqref{eq:bh_mixing} and \eqref{eq:bmass}, to obtain
\begin{align}\label{eq:theta_mb_muB>>mB}
&\theta_{hb}\simeq -\frac{\mubv v}{m_h^2}\sin(\alphabvp) \simeq \frac{f_B M \mbv(|\xi_2|^2+|\xi_3|^2)}{8\pi^2 v m_h^2}\sin(\deltabv),\\
& m_b^2 \simeq \frac{\mubv v^2}{2f_B}\cos(\alphabvp)\simeq \frac{\mubv v^2}{2f_B}.\nonumber
\end{align}
It is important to emphasise that the equation above is valid once the two terms in Eq.\,\eqref{eq:b_tadpole} are comparable:
\begin{equation}\label{eq:muB_mB}
\frac{\mubv}{\mbv}\gtrsim \frac{ M (|\xi_2|^2+|\xi_3|^2)}{4\pi^2v^2}f_B.
\end{equation}
Then, combining Eqs.\,\eqref{eq:theta_mb_muB>>mB} and \eqref{eq:muB_mB} leads to
\begin{equation}\label{eq:theta_mass_scaling}
\frac{\theta_{hb}}{m_b^2} \simeq \frac{f_B\sin(\deltabv)}{m_h^2v}, \quad (\mubv\gg \mbv),
\end{equation}
that shows $\theta_{hb}\propto m_b^2$, with the proportionality constant depending linearly on $f_B\sin(\deltabv)$.

\section{UV completion}\label{app:UV}

In this appendix we discuss possible tree-level UV completions that generate at low energies the effective interactions present in Eqs.\,\eqref{eq:Leff_noflavor} and \eqref{eq:Leff}. The new particle should have at least two new interactions - there should be a coupling to two quarks and another to a quark and $N$. Because of this, we are left to consider diquark type UV completions, i.e. particles that couple to the $\Delta B = 2/3$ bilinears $u u$, $d d$ or $u d$. Since we are interested in the neutral $udd$ combination, we do not consider bilinears with two flavors of up-quarks. Writing the possible chiralities, the diquark can couple to the following operators
\begin{equation}
    d_R d_R,~ u_R d_R,~ d_R u_R,~ \big(d_L d_L,~ u_L d_L,~ d_L u_L\big) \subset Q_L Q_L.
\end{equation}

\noindent The other coupling between the diquark, one quark and one neutral fermion $N$ allows to generate the desired effective operator. This $NXq$ coupling fixes the representation of the diquark $X$ to be a $\mathbf{3}$ of $SU(3)_c$, since $Xq$ should be color neutral. A similar argument holds for $SU(2)_L$; since $Xq$ should be $SU(2)_L$ neutral, $X$ must be either a singlet in the case of $q=u_R, d_R$ or a doublet in the case of $q=Q_L$. However, if $X$ is an $SU(2)_L$ doublet, it cannot couple to any quark bilinear, so we shall not consider this case. Lastly, we are left with two possibilities for the hypercharge. If $X$ couples to $N u_R$, it must have hyperchage $2/3$. Alternatively, if $X$ couples to $Nd_R$ it must have hypercharge $-1/3$. Therefore, we can write the desired operators
\begin{align}
    \label{eq:X2/3}&
    X ~d_R d_R + X^\dagger N u_R +h.c.
    \\
    \label{eq:X1/3} \text{or} \quad &
    \widetilde{X}^\dagger ~Q_L Q_L + \widetilde{X}^\dagger ~u_R d_R + \widetilde{X} N d_R +h.c.,
\end{align}

\noindent where $X = (\mathbf{3},\mathbf{1})_{2/3}$ and $\widetilde{X} = (\mathbf{3},\mathbf{1})_{-1/3}$. Notice that the UV completions where the diquark couple to $Q_L Q_L$ and $u_R d_R, d_R u_R$ are both the one with $\widetilde{X}$. While, the interactions of $X$ are only the ones given in Eq.\,\eqref{eq:X2/3}, for $\widetilde{X}$ we can write couplings to leptons too. Restoring the four-component notation, the full tree-level Lagrangian for both terms are
\begin{alignat}{5}
    \label{eq:L2/3}&
    \mathcal{L}_{X}&&= \lambda_{\alpha i} X^\dagger \overline{N^c_\alpha} u_R^i + \lambda'_{jk} X ~\overline{d_R^c}^j d_R^k + h.c. ,
    \\ \nonumber  &
    \mathcal{L}_{\tilde X}&&= \eta_{\alpha i} \widetilde{X} \overline{N_\alpha^c} d_R + \eta'_{jk} \widetilde{X}^\dagger \overline{Q_L^c}^j Q_L^k + \eta''_{jk} \widetilde{X}^\dagger \overline{u_R^c}^j d_R^k 
    \\ \label{eq:L1/3} &
    &&+ \eta'''_{jk} \widetilde{X} \overline{u_R^c}^j e_R^k + \eta''''_{jk} \widetilde{X} \overline{Q_L^c}^j L_L^k + h.c.
\end{alignat}

\noindent Due to the $\eta'''$ and $\eta''''$ couplings to leptons in the case of $\widetilde{X}$, the proton decays at tree-level. Since the bounds on proton decay for diquarks at the $\TeV$ scale are very strong\footnote{To keep the $\widetilde{X}$ theory, one would have to tune the product of the couplings $\eta''' \eta''''$ to be below $10^{-19}$\,\cite{Alonso-Alvarez:2023bat}. We discard this possibility since the alternative, $X$, does not have this issue.}, from now on we will only consider  the UV completion involving $X$.

The UV completion with a scalar diquark $X$ has some interesting features. The immediate concern for baryon number violation at energies around the weak scale is proton decay. However, it is possible to have $B$ violation without proton decay, as it was explored in several occasions in the literature before\,\cite{Giudice:2011ak,Arnold:2012sd,Assad:2017iib,Cheung:2013hza,Davoudiasl:2010am}. The hypercharge $2/3$ diquark model does not induce proton decay provided that the majorana fermions $N$ do not mix with the SM neutrinos and are heavier than $\sim 1\GeV$. Regarding the flavor structure of the interactions, color antisymmetry in Eq.\,\eqref{eq:L2/3}, means that $\lambda'_{jk}$ must be antisymmetric in flavor. This property is preserved after defining the quark mass basis in the electroweak broken phase since the Cabibbo--Kobayashi-Maskawa (CKM) matrix transformation is the same for the two $d_R$ fields. This flavor structure implies that there are three independent $\lambda'$ couplings, 
\begin{equation}
    \label{eq:lambprime}\lambda'_{jk} =\epsilon_{jkl}\lambda'_l.
\end{equation}

\noindent Because of the antisymmetric nature of the diquark interactions, there are no tree-level $\Delta F = 2$ processes. Additionally, any one-loop meson mixing must involve all three generation of quarks. For example, for kaon oscillations, the diagrams must involve the $b$ quark and for $B$-meson mixing the loop must contain an $s$ quark. So, if one of the couplings $\lambda'_{s,b}\ll \lambda'_{ds},\lambda'_{db}$ is small, meson oscillations can be suppressed while still allowing for order one diquark couplings in the other flavors. Therefore, the bounds from $K^0\bar K^0$ and $B^0\bar B^0$ mixing are naturally suppressed. Considering the hierarchies in the CKM sector, such hierachical diquark structure is not unreasonable and may appear in many UV scenarios as argued in Ref.\,\cite{Giudice:2011ak}.

\bibliography{Bgen.bib}
\bibliographystyle{bibi}

\end{document}